\documentclass[aps,prl,twocolumn,superscriptaddress]{revtex4}
\usepackage[sort&compress]{natbib} 
\usepackage{graphicx}   
\usepackage{epstopdf}
\usepackage{dcolumn}
\usepackage{bm}
\usepackage{amsmath}
\usepackage{booktabs}
\usepackage{multirow}
\usepackage{gensymb}
\usepackage{color}
\usepackage{units}

\renewcommand{\vec}[1]{\bm{#1}}

\begin{document}
\title{ Magnetoelectro-elastic control of magnetism in an artificial multiferroic}

 \author{J. Heidler}
 \affiliation{Swiss Light Source, Paul Scherrer Institut, 5232 Villigen PSI, Switzerland}
 \affiliation{SwissFEL, Paul Scherrer Institut, 5232 Villigen PSI, Switzerland}

\author{M. Fechner}
 \affiliation{Materials Theory, ETH Z\"{u}rich, 8093 Z\"{u}rich, Switzerland}

 \author{R. V. Chopdekar}
\email[Author to whom correspondence should be addressed. Email: ]{rchopdekar@ucdavis.edu}%
\affiliation{Swiss Light Source, Paul Scherrer Institut, 5232 Villigen PSI, Switzerland}
 \affiliation{Laboratory for Micro- and Nanotechnology, Paul Scherrer Institut, 5232 Villigen, Switzerland} 
 \affiliation{Department of Chemical Engineering and Materials Science, University of California-Davis, Davis, CA 95616, USA} 
   
\author{C. Piamonteze}
\email[Author to whom correspondence should be addressed. Email: ]{cinthia.piamonteze@psi.ch}%
 \affiliation{Swiss Light Source, Paul Scherrer Institut, 5232 Villigen PSI, Switzerland}

 \author{J. Dreiser}
\affiliation{Swiss Light Source, Paul Scherrer Institut, 5232 Villigen PSI, Switzerland}

  \author{C. A. Jenkins}
 \affiliation{Advanced Light Source, Lawrence Berkeley National Laboratory, Berkeley CA 94720, USA} 
\author{E. Arenholz}
\affiliation{Advanced Light Source, Lawrence Berkeley National Laboratory, Berkeley CA 94720, USA} 
 
   \author{S. Rusponi}
 \affiliation{\'{E}cole Polytechnique F\'{e}d\'{e}rale de Lausanne, Institute of Condensed Matter Physics, 1015 Lausanne, Switzerland}

 \author{H. Brune}
\affiliation{\'{E}cole Polytechnique F\'{e}d\'{e}rale de Lausanne, Institute of Condensed Matter Physics, 1015 Lausanne, Switzerland}
  
  \author{N. A. Spaldin}
\affiliation{Materials Theory, ETH Z\"{u}rich, 8093 Z\"{u}rich, Switzerland}

  \author{F. Nolting}
  \affiliation{Swiss Light Source, Paul Scherrer Institut, 5232 Villigen PSI, Switzerland}

\date{01.09.2015}

\begin{abstract}
 
We study the coexistence of strain- and charge-mediated magnetoelectric coupling in a cobalt ($0-7$~nm) wedge on ferroelectric [Pb(Mg$_{1/3}$/Nb$_{2/3}$)O$_{3}$]$_{0.68}$-[PbTiO$_{3}$]$_{0.32}$ (011) using surface-sensitive x-ray magnetic circular dichroism spectroscopy at the Co L$_{3,2}$ edges. Three distinct electric field driven remanent magnetization states can be set in the Co film at room temperature. 
\emph{Ab-initio} density functional theory calculations unravel the relative contributions of both strain and charge to the observed magnetic anisotropy changes illustrating magnetoelectro-elastic coupling at artificial multiferroic interfaces. 
 \end{abstract}
  
\maketitle
Multiferroic systems, where two or more ferroic properties (ferromagnetism, ferroelectricity or ferroelasticity) coexist, provide the opportunity to study coupling mechanisms between different order parameters \cite{Ramesh07}. The prospect of electric field control of magnetism with its potential use in technological applications \cite{Mathews1997,Gajek2007,Burton2011,Hu11} has focused attention on the subgroup of multiferroics that exhibits magnetoelectric (ME) coupling.  Heterostructures consisting of cross-coupled ferromagnetic (FM) and ferroelectric (FE) layers \cite{Vaz12} are often referred to as artificial multiferroic composites. Due to their modular nature the number of available systems with potential multiferroic properties is greatly increased compared to intrinsic multiferroic systems which proves to be an advantageous concept to achieve electric field control of magnetism at room temperature. 

The mechanisms involved in ME interface coupling often result from new and interesting underlying physical phenomena. 
Strain-coupled systems  \cite{Eerenstein2007, Zhang12CoFeB, WuPMNPT} make use of the piezoelectric properties of a ferroelectric system to control the magnetism in a ferromagnet through magnetostriction.
Furthermore, ferroelectric polarization reversal may change the overlap between atomic orbitals at the FM/FE interface \cite{DuanPRL2006,BurtonLSMO09,Valencia2011} or drive a magnetic reconstruction \cite{Borisov2005, Bea2008, Molegraaf09, VazPZTLSMO2010, He2010, Wu2010} at the interface. 
Charge-mediated ME coupling exploits the electric field effect \cite{ Zhang99, Duan2008, Rondinelli08, marujama09} as well as the remanent electric polarization of FE components \cite{Zhao2005, Stolichnov2008} to modulate the charge carrier concentration in an adjacent FM layer, where accumulation or depletion of spin-polarized electrons results in a change of the interface magnetization. 
Different length scales apply to the aforementioned mechanisms. While 
 the influence of strain extends to several tens of nanometers, the screening of surface charge takes place within the Thomas-Fermi screening length (on the order of a few Angstroms in metals \cite{Zhang99}). 
The coexistence of strain and charge effects have seldom been reported \cite{NOBTOShu2012,Hustrainandcharge11,NanPMNPT,Pertsev15}  and so far been explained in a  phenomenological framework. 
In this work, we disentangle strain and charge contributions to the magnetic response upon electrical switching using surface-sensitive x-ray magnetic circular dichroism (XMCD) at the Co L$_{3,2}$ edges and \emph{ab-initio} density functional theory (DFT). The heterostructure consisting of a Co wedge (0-7~nm) grown on top of the ferroelectric [Pb(Mg$_{1/3}$Nb$_{2/3}$)O$_{3}$]$_{0.68}$ -[PbTiO$_{3}$]$_{0.32}$ (011) (from here on PMN-PT) allows for a thickness-dependent study. We find experimentally that it is possible to set three distinct remanent and reversible magnetization states through \emph{magnetoelectro-elastic} control at room temperature. DFT calculations for different strain and charge states reproduce the experimental behavior and unravel the different mechanistic contributions.

\begin{figure} [t]

\includegraphics[width=0.45\textwidth]{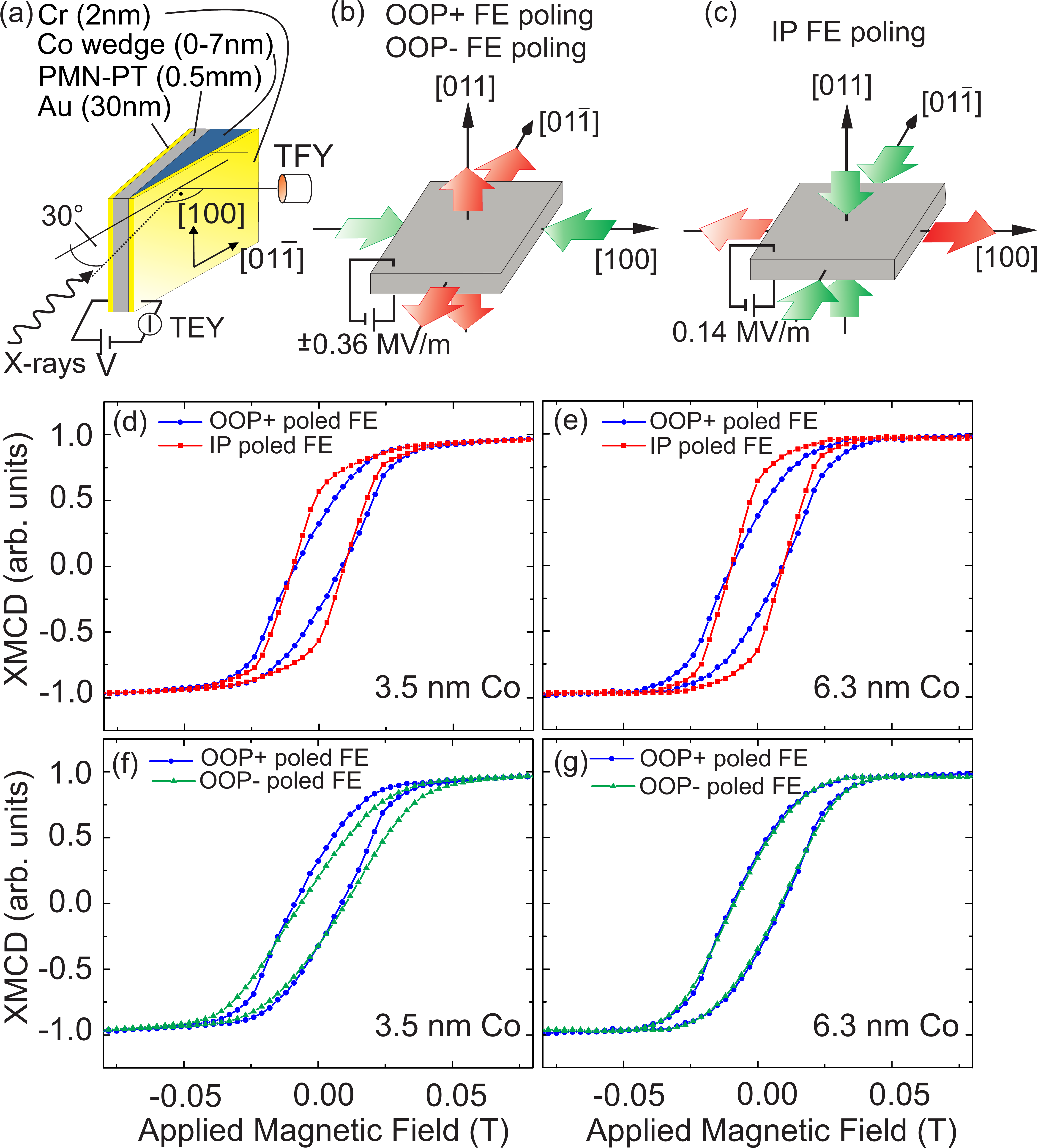}
 \caption{(a) Measurement geometry. (b,c) Lattice parameter changes in OOP/IP poled PMN-PT, respectively. Green (red) arrows indicate compressive (tensile) strain in Co. (d-g) XMCD hysteresis curves probing the Co magnetization projection along the ($01\overline{1}$) PMN-PT crystal direction 
  for the three distinct FE polarization states. (d,e) Switching the FE polarization from an OOP (blue curve) to an IP poled state (red curve) probing a nominal Co thickness of 3.5~nm  (d) and 6.3~nm (e), induces an anisotropy change with higher remanent magnetization. 
(f) For 3.5~nm Co thickness, OOP poled polarization directions exhibit also different anisotropies. (g) For 6.3~nm Co thickness the anisotropy change for oppositely OOP poled FE is now absent. } \label{hyst}
  \end{figure}

Relaxor FE PMN$_{(1-x)}$-PT$_{x}$ (011), with a composition  of $x=0.32$ located in the morphotropic phase boundary region \cite{Noheda02}, (Atom Optics Co., LTD., Shanghai, China) is used as a substrate due to its strong piezoelectric properties. Its crystal structure is monoclinic with lattice constants $a$=4.02~\r{A} , $b$=4.01~\r{A} and $c$=4.03~\r{A} \cite{Noheda02}. A cobalt wedge with linearly increasing thickness from $0-7$~nm is grown on PMN-PT (011) via thermal evaporation. X-ray diffraction showed that Co grows  with fcc (111)-texture. 
 A capping of 2~nm Cr prevents oxidation and a 30~nm Au film serves as bottom electrode. Figure~\ref{hyst}~(a) shows the sample design and measurement geometry.
Depending on the electric field applied across PMN-PT (011), three distinct remanent FE polarization states can be set.
 The FE polarization is poled positively or negatively out of plane (OOP+ or OOP-) by applying an electric field of $\pm0.36$~MV/m at the bottom electrode while the top electrode is connected to ground.  
 When comparing OOP+ and OOP- poled FE no lattice parameter change in PMN-PT is expected and the Co top layer encounters identical strain conditions. However, FE polarization switching alters the interfacial charge that has to be screened by the adjacent cobalt layer through accumulation or depletion of electrons. 
Sweeping between opposite OOP FE polarization directions, PMN-PT (011) exhibits a remanent in-plane (IP) poled state at the coercive electric field ($\pm0.14$~MV/m). 
 The switching from an OOP to an IP poled configuration and vice versa is accompanied by structural changes of the PMN-PT \cite{strainnote,mypaper} as indicated in Fig.~\ref{hyst}~(b) and (c) that act on the Co top layer. OOP$\leftrightarrow$IP switching alters both the strained state of cobalt and the interfacial charge seen by the Co film. Note that both OOP poled states as well as the IP poled configuration are stable at remanence. The FE polarization of PMN-PT (011) at 298~K was measured to be $2\cdot\vec{P}_{\textnormal{PMN-PT}}=$\unit[60]{$\mu$C/cm$^2$}.

XMCD \cite{Stohr06} measurements at the Co $L_{3,2}$ edges were carried out at the X-Treme beamline \cite{Cinthiabeamline} at the Swiss Light Source, Paul Scherrer Institut, Switzerland and at beamline 6.3.1  \cite{ALS} at the Advanced Light Source, Lawrence Berkeley National Laboratory, California, USA. In XMCD, the absorption intensity difference between opposite light helicities is an element sensitive probe of magnetization along the photon propagation direction. Spectra were recorded at room temperature in grazing incidence geometry, where x-rays are inclined from the surface plane by an angle of 30\degree, measuring the projected magnetization along the ($01\overline{1}$) crystal direction of the PMN-PT. The external magnetic
field was applied along the x-ray beam direction. Sum rules allow for a quantification of the Co spin and orbital magnetic moments $m_\mathrm{s,eff}$ and $m_\mathrm{orb}$ from analysis of the XMCD spectra \cite{Carra1993, Zhengsumrule95}.

\begin{figure} [t]
\includegraphics[width=.38\textwidth]{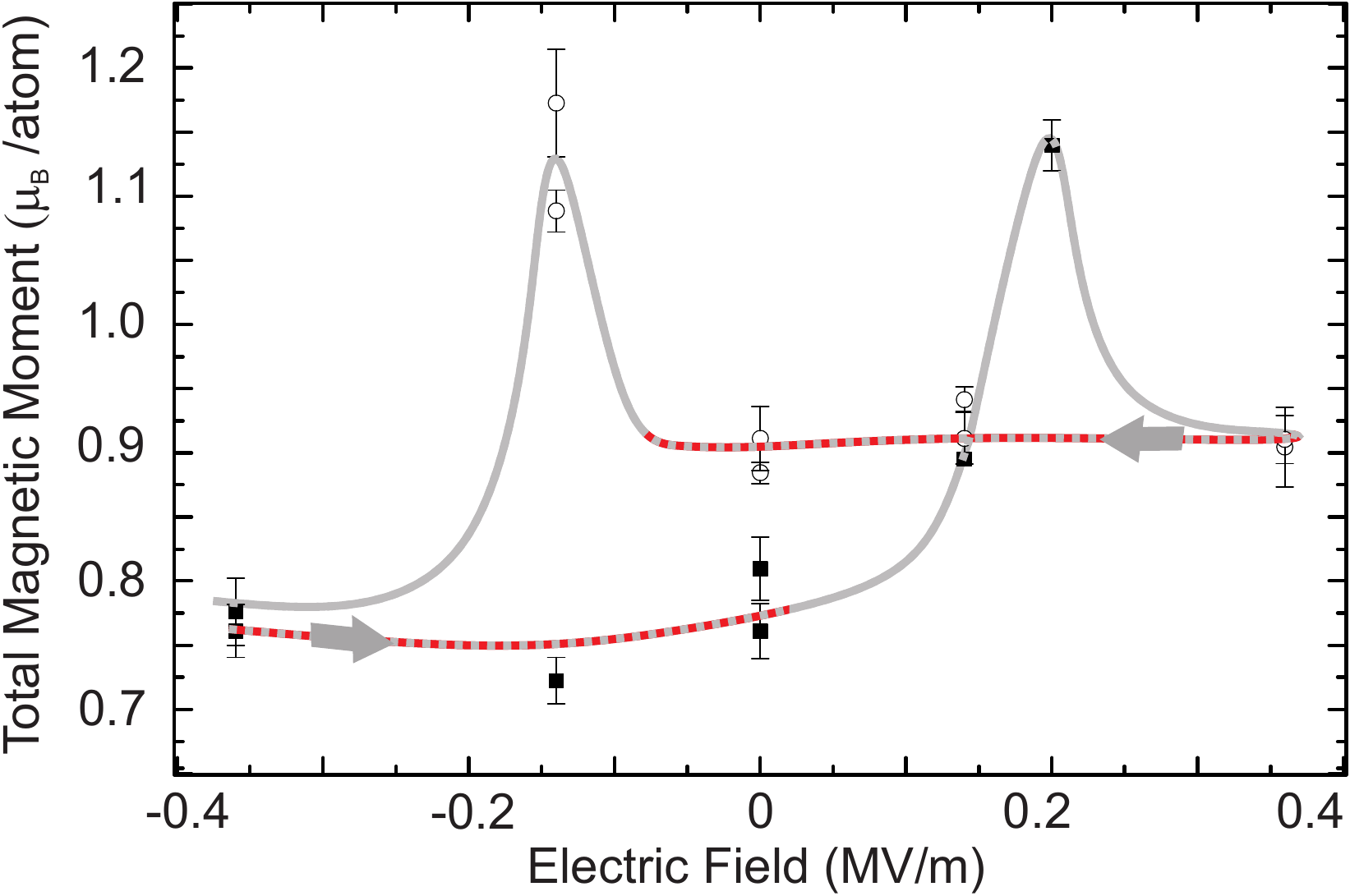}
\caption{ Total magnetic moment along the (${01\overline{1}}$) direction at remanence obtained from sum rule analysis as a function of applied electric field. The grey curve is a guide for the eye to link subsequent measurements. The arrow indicates the sweeping direction. Black squares (open circles) indicate measurements coming from OOP- (OOP+) poling. The dashed red branches highlight the dependence of $m_\mathrm{tot}$ on the FE OOP substrate polarity  when switching between OOP+ and OOP-.} \label{spinmoment}
\end{figure}
   
Co XMCD hysteresis loops along the ($01\overline{1}$) direction, taken in  total electron yield (TEY) mode for oppositely OOP poled states as well as the IP poled configuration at distinct thicknesses of the wedge, highlight two different ME coupling mechanisms at play. Electrical switching from an OOP poled to an IP poled state induces an anisotropy change with higher remanent magnetization as seen in Fig.~\ref{hyst}~(d) for a nominal Co thickness of 3.5~nm. The same behavior is observed probing a thicker part of the wedge at a nominal Co thickness of 6.3~nm  in Fig.~\ref{hyst} (e). Additionally we observe a more subtle anisotropy change comparing hysteresis curves taken for oppositely OOP poled FE in Fig.~\ref{hyst}~(f). This anisotropy change is not observed in the thicker part of the wedge, as seen by the nearly identical hysteresis loops of  Fig.~\ref{hyst}~(g).
 TEY is a surface-sensitive detection mode where the  probability of electron escape  from the Co/PMN-PT interface decays exponentially with increasing Co top layer thickness (the electron sampling depth for Co is about 2.5~nm ~\cite{Nakajima99}). 
Therefore, the observed difference in magnetic anisotropy in Fig.~\ref{hyst}~(f) and its absence in Fig.~\ref{hyst}~(g) hints that its origin lies at the interface between Co and PMN-PT. As pointed out above, this effect cannot be attributed to a piezoelectric-magnetostrictive coupling since the structure of PMN-PT in the two states is equivalent. Hence, this anisotropy change due to the substrates' opposite OOP polarities suggests a charge driven magnetoelectric coupling.
 The anisotropy change shown in Fig.~\ref{hyst} (d) and Fig.~\ref{hyst} (e) at both the thinner and the thicker part  of the wedge can be understood in terms of the magnetostriction of cobalt in response to the lattice parameter changes of PMN-PT \cite{mypaper}. Since strain is a `{bulk}' effect,  its influence persists throughout the whole Co film thickness.
For a quantitative analysis, a series of XMCD spectra was taken as a function of applied electric field on the thin part of the wedge at 3.5~nm Co thickness at magnetic remanence after saturation in 2~T in total fluorescence yield (TFY).  
Sum rule analysis was used to extract the magnetic moment  $m_\mathrm{tot}=m_\mathrm{{s,eff}}+m_\mathrm{orb}$ projected along the (${01\overline{1}}$) direction (for details, see `Supplemental Material'). The resulting dependence on the electric field is given in Fig.~\ref{spinmoment}, where the gray curve links successive measurements. $m_\mathrm{tot}$ is strongest at the coercive electric field, where the FE polarization is rotated in-plane. Comparing measurements of oppositely poled FE, OOP- poled PMN-PT results in a smaller Co $m_\mathrm{tot}$ than OOP+ poled PMN-PT. Here, $m_\mathrm{tot}$ depends solely on the FE polarization state that the PMN-PT has been set in, irrespective of an actively applied bias voltage.  
Note that in 2~T applied field no dependence of the saturation magnetization on the FE polarization can be observed. At 2~T  field applied along the easy (100) direction, the effective spin moment $m_\mathrm{{s,eff}}=1.64 \pm 0.16~\mu_\mathrm{B}$ and orbital moment $m_\mathrm{orb}= 0.131  \pm 0.002~\mu_\mathrm{B}$ compare well with literature values \cite{Zhengsumrule95, Tischer95}. 

The impact of the FE order of PMN-PT on the electronic and atomic structure of a Co top layer is twofold.
We observe a hysteretic behavior of remanent $m_\mathrm{tot}$ for OOP+ and OOP- poled FE suggesting a charge-driven magnetoelectric coupling contribution due to accumulation and depletion of electrons at the FM/FE interface. 
The contribution of charge to the change in total magnetic moment is highlighted by the dashed red branches in Fig.~\ref{spinmoment}. 
 Deviations occur only at the coercive electric field, where strain dominates while no net surface charge should be present. 
As the total moment at 2 T does not appreciably change with FE polarization but there is a significant change to $m_\mathrm{tot}$ at magnetic remanence, we attribute these changes in magnetization to changes in effective magnetic anisotropy energy (MAE) of the Co film.
To investigate the separate influences of strain and screening charge on the MAE we perform first-principles DFT calculations of bulk fcc cobalt with each perturbation applied separately. 

 For Co films thicker than 1.5~nm \cite{Sander:1999vp,Bruno:1988wn,Chappert:1988fa}, the shape anisotropy dominates the MAE and dictates an isotropic in-plane magnetization. 
This isotropy within the film plane is subsequently lifted by other MAE contributions. The bulk magnetocrystalline anisotropy for fcc Co favors  an easy axis along the [111]- and equivalent cubic directions. However, for a (111) film the strong shape anisotropy disfavors the low energy crystalline directions. Moreover, the volume magnetocrystalline anisotropy is isotropic within the (111) film plane and thus creates no anisotropy even if its magnitude is altered.

Another contribution to the MAE is magnetoelasticity, which exhibits lower order terms of the directional magnetization expansion \cite{Sander:1999vp} that are coupled to strain tensor elements ($\epsilon_{ij}$). For cubic symmetry its energy contribution is  
\begin{eqnarray}
E_{mag-el}&=&B_1 (\epsilon_{11}\alpha_1^2+2\epsilon_{22}\alpha_2^2+\epsilon_{33}\alpha_3^2)\nonumber\\
& &+2B_2(\epsilon_{23}\alpha_2\alpha_3+\epsilon_{13}\alpha_1\alpha_3+\epsilon_{12}\alpha_1\alpha_2) \nonumber \;,
\end{eqnarray}
where $B_i$ are the cubic magnetoelastic constants and $\vec{\alpha}$ is the corresponding direction cosine of the magnetization. For the [111]-oriented fcc Co film we transform this expression \cite{Sander:1999vp} (see `Supplemental Material') into hexagonal coordinates to yield, for the film plane magnetization:
\begin{equation}
E_{mag-el,hex}(\phi)=-\frac{1}{3}(B_1+2 B_2) (\epsilon'_{100}-\epsilon'_{01\overline{1}}) \sin ^2(\phi )\;
\end{equation} 
where $\epsilon'_{i}$ are the strain elements in the film-plane labeled with respect to the PMN-PT substrate and $\phi$  is the angle of the in-plane magnetization relative to the [100] direction. The magnetoelasticity creates an easy in-plane direction which is determined by an `effective' magnetoelastic constant $B_{\textnormal{eff}}=B_1+2 B_2$. 

By performing total energy calculations for a set of strained fcc-cobalt unit cells (see `Supplemental Material') we compute $B_1$ and $B_2$ using DFT. We find both $B_{1}$=$-8.7~\unit{MJ m^{-3}}$ and $B_2$=$7.2~\unit{MJ m^{-3}}$ in reasonable agreement with experimental and theoretical literature values \cite{Anonymous:6Y_8yj2R,Sander:1999vp,Guo:1999hr}. Moreover, the combination of these values gives a positive effective magnetoelastic constant,  $B_{\textnormal{eff}}$. Consequently, we predict that a net strain ($\epsilon'_{100}-\epsilon'_{01\overline{1}})>0$ creates an easy axis along the $[01\overline{1}]$ direction, whereas ($\epsilon'_{100}-\epsilon'_{01\overline{1}})<0$ will produce an easy axis parallel to $[100]$. In PMN-PT,  OOP$\rightarrow$IP poling is accompanied by a strong positive $\epsilon'_{100}$ transferred to the Co film \cite{strainnote} resulting in a positive net strain. Hence, our theoretical finding is in agreement with the experimentally observed anisotropy change along $[01\overline{1}]$  upon IP poling. 

For both the OOP+ and the OOP- poled state, the Co film encounters a net strain ($\epsilon'_{100}-\epsilon'_{01\overline{1}})<0$ and the experimentally observed magnetization shows a preferred orientation close to the $[100]$ axis in agreement with our prediction. However, in the experiment there is a 15\% higher magnetization projection along the $[01\overline{1}]$ axis for the OOP+ state than for the OOP- state. Since the structure of PMN-PT in the two states is equivalent, the difference has to be attributed to a contribution stemming from the FE polarization direction.

For example, the presence of interface charge $\sigma_{int}$ may necessitate screening by the valence electrons of the adjacent Co film.
 With $2\cdot\vec{P}_{\textnormal{PMN-PT}}$=\unit[60]{$\mu$C/cm$^2$}, the amount of interface charge doping for fcc (111) Co 
can be estimated to be $\sigma_{int}(0) =\pm$\unit[0.102]{~e$^{-}$/unit cell area}. 
 This charging will be largest at the interface and then decay exponentially corresponding to the Thomas-Fermi screening as $\sigma_{int}(z)=\sigma_{int}(0) e^{-z/\lambda_\textnormal{Co}}$, where $z$ measures the distance from the interface and $\lambda_\textnormal{Co}$ is the Thomas-Fermi-screening length of Co ($\lambda_\textnormal{Co}=$\unit[0.15]{nm} \cite{Zhang99}). 

\begin{figure} [t]
\includegraphics[width=.45\textwidth]{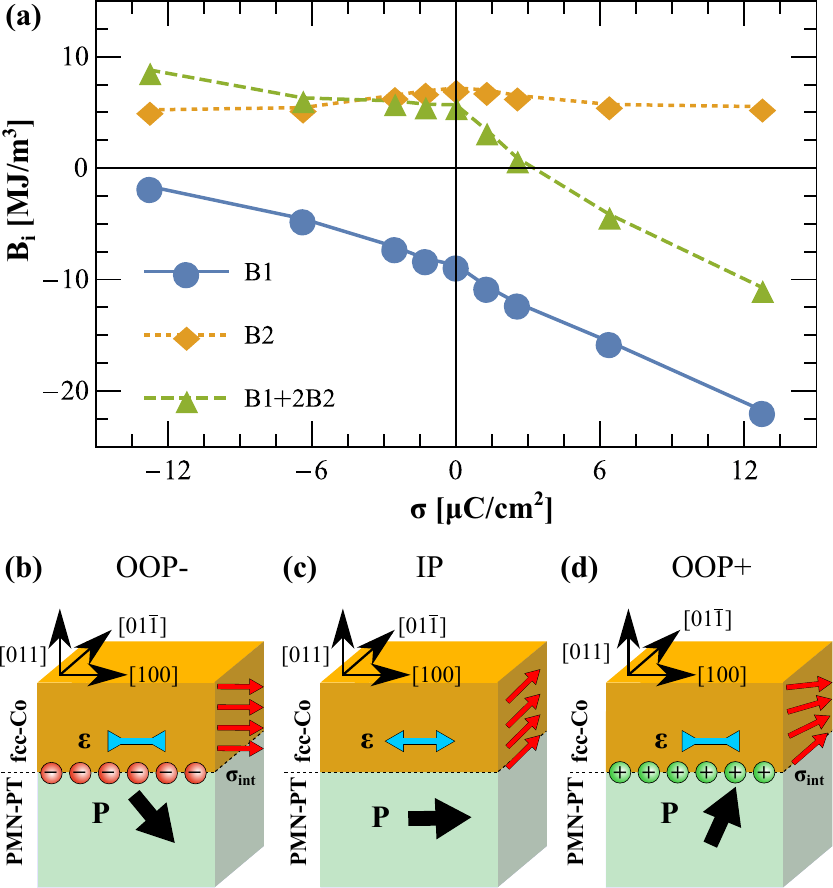}
\caption{(a) Variation of magnetoelastic constants as a function of charging of the unit cell $\sigma$. (b,c,d) Sketch  of the three switching states (OOP-, IP, OOP+) of the PMN-PT/Co interface. The black/red arrows show the direction of electric/magnetic polarization and the blue arrows show the net strain. (b) illustrates the OOP- case where strain and electron charge lead to a preference of the [100] direction as the easy axis. (c) depicts the highly strained IP state with no interface charging and an easy axis along  [$01\overline{1}$].  In (d) the combination of strain and positive interface charge creates an easy axis along  [$01\overline{1}$] at the interface which decays and turns towards [100] away from the interface.}\label{theory}
\end{figure}

Next we examine the impact of this interface charge on the magnetoelastic constants ($B_1(\sigma)$, $B_2 (\sigma)$), as shown in Fig.~\ref{theory}(a), by repeating our computations with a varied total $e^{-}$ count within the DFT calculations. We find (Fig.~\ref{theory}(a)) a strong variation of $B_1$ with charging whereas $B_2$ remains nearly unchanged. Moreover, the different behavior of $B_1$ and $B_2$ as a function of charging leads to a sign change of $B_{\textnormal{eff}}$ around $\sigma=$\unit[4]{$\mu C/cm^2$}, as depicted by a dashed green line in Fig.~\ref{theory}(a). For the same negative net strain corresponding to OOP poled PMN-PT, the OOP+ and OOP- cases have different alignments of the magnetic easy axis at the interface, as sketched in Fig.~\ref{theory}~(b,d). For the OOP- case, the accumulation of positive charges at the interface ($\sigma<0$, $B_{\textnormal{eff}}>0$) creates an easy axis along the $[100]$ direction. On the other hand, in the OOP+ state the accumulation of negative charges ($\sigma>0$) reverses the sign of $B_{\textnormal{eff}}$ and thus favors the orthogonal $[01\overline{1}]$ direction as easy magnetization direction. Consequently, we expect that switching of the electric polarization in combination with an alternation of the magnetoelastic constants by interface charging leads to a 90\degree change of the preferred magnetization direction. This \emph{magnetoelectro-elastic} effect will be constrained to the interface region, where enough charge accumulation is present.

The experimentally observed  higher remanent magnetization along the [$01\overline{1}$] direction for the OOP+ state  compared to the  OOP- state in Fig.~\ref{hyst} and \ref{spinmoment} reflects our calculated \emph{magnetoelectro-elastic} effect. However,  the detected signal contains contributions of both, the strain that extends throughout the entire film, as well as the charge --- an interface effect. Consequently, the exponential decay of the charge screening away from the Co/PMN-PT interface suppresses the measurement of the predicted effect  in thicker films when using surface-sensitive TEY detection mode (Fig.~\ref{hyst}(g)). 

In conclusion, we investigated by a combined experimental and theoretical effort the magnetic properties of the artificial multiferroic Co/PMN-PT interface. From our XMCD measurements we found that the magnetic anisotropy behavior of the Co film depends on the three distinct polarization states (IP, OOP(+,-)) the PMN-PT can be set in.   For thin film thicknesses in which interface effects dominate we find a significant difference between all three states, whereas for thick film thicknesses the difference between the OOP states vanishes. Our theoretical investigation illustrates that the changes in anisotropy are due to a combination of magnetoelasticity and interface charging opening up the possibility for enhanced magnetoelectric coupling. Finally, we suggest that the found modulation of magnetic anisotropy by the \emph{magnetoelectro-elastic} effect may allow to create a magnetic anisotropy gradient in thin films. If the gradient is strong enough, it could give rise to a spiral state in the thin film, which could be controlled by the ferroelectric substrate.

\begin{acknowledgments}
This work was supported by the Swiss Nanoscience Institute and EU's 7th Framework Programm IFOX (NMP3-LA-2010 246102). The x-ray absorption measurements were performed on the EPFL/PSI X-Treme beamline at the Swiss Light Source, Paul Scherrer Institut, Villigen, Switzerland and at beamline 6.3.1 at the Advanced Light Source, Lawrence Berkeley National Laboratory, California, USA. 'The Advanced Light Source is supported by the Director, Office of Science, Office of Basic Energy Sciences, of the U.S. Department of Energy under Contract No. DE-AC02-05CH11231. We thank Christof Schneider for his assistance in structural characterization and Marcus Schmidt for technical support. 
\end{acknowledgments}
 

\begin{thebibliography}{43}
\bibitem[{str()}]{strainnote}
\bibinfo{note}{X-ray diffraction reciprocal space maps on PMN-PT (011) show
  that switching to an IP poled state is accompanied by a tensile lattice
  parameter change of $+0.90$~\% along the (100) direction, as well as
  compressive lattice parameter changes of $-0.23$~\% along the
  ($01\overline{1}$) and $-0.70$~\% along the (011) direction, exerting strain
  to the Co layer.}
\end{thebibliography}

\end{document}